\documentclass[doublespacing]{elsart}
\usepackage[dvips]{graphicx}
\usepackage{epsfig}
\usepackage{natbib}
\usepackage{amssymb}
\begin{document}
\begin{frontmatter}
\title{The role of mutation rate in a simple colonization model}
\author{R. Huerta-Quintanilla\corauthref{qw}}
\ead{rhuerta@mda.cinvestav.mx}
\author{and M. Rodr\'{\i}guez-Achach}
\address{\it Departamento de F\'{\i}sica Aplicada Unidad M\'erida, 
Cinvestav-IPN,
Km.\ 6 Carretera Antigua a Progreso, M\'erida 97310, Yucat\'an, M\'exico.}
\corauth[qw]{Corresponding author}
\begin{abstract}
We study the effect of mutations in a simple model of colonization, 
based on Montecarlo simulations. When the population
colonizes the whole available habitat, a maximum population density
is reached, which depends on the mutation rate. Depending
on the values of other parameters, such as selection pressure,
fecundity and mobility, there is an optimal value for 
the mutation rate for which the colonization reaches
the highest density. We also investigate the survival probabilities
under different conditions and its relation to the mutation rate.
\end{abstract}

\end{frontmatter}

\section{Introduction}

Modeling biological evolution has a long tradition among
biologists and applied mathematicians. More recently, the
physics community has been attracted to this area because
of the rich and complex behavior that arises from these
systems, and the possibility to use the available tools from
statistical mechanics and computer simulations.

There are many interesting questions that one encounters
when dealing with the evolution of populations. Under what
conditions can a population successfully colonize a new
habitat? What are the factors that most strongly influence this process?
Why species do not spread indefinitely into a new territory?
Some of the factors that have been studied are the accumulation
of harmful mutations \cite{lande,lynch,penna}, selection
pressure \cite{burger,holt} and changing environment
\cite{pek,kirk}, among others.

In this paper we use a model very similar to that recently
introduced by P\c ekalski \cite{pek2}. We study the process
of colonization of a habitat by a
population initially living on a small region of the
habitat. We consider the climate to be spatially
varying through this habitat. The individuals are characterized
by a single feature, their phenotype, which determines
the survival chances by comparison with the ideal
conditions imposed by the habitat climate. In this work
we are primarily interested in the role played by
mutations in the colonization process. We investigate
the optimal rate of mutation and its dependence
on several model parameters : fecundity, selection pressure, 
and mobility.

\section{Model}

The habitat is considered to be a lattice of size $x=200 \times y=100$.
At most one individual can occupy a lattice site at a given
time. The climate is defined by assigning to each lattice site
a real number $F(x)$ between 0 and 1. The variation is linear
from $x=0$ to $x=200$. The adaptation or 
fitness of a given individual
to the environment can be measured by the difference between
its phenotype, which is also a single real number $z$ between
0 and 1,  and the ideal one, $F$ \cite{mroz,frazer,morgan}.
Following P\c ekalski \cite{pek2}, the fitness of an
individual $i$ at position $x_i$ is calculated as
$$
p_i=e^{-\alpha|z_i-F(x_i)|},
$$
where $\alpha$ is the selection pressure. The climate $F$
varies from 0 on the leftmost part of the habitat ($x=0$) to  a  value
of 1 to the right. Individuals are hermaphrodites, thus the only condition
for mating is that they are nearest neighbors. Effects of 
inbreeding are also neglected in this model. The parents die
after reproduce. We have not implemented an aging mechanism to
keep the model as simple as possible.

The simulation proceeds as follows:

1. An individual $i$ is picked at random.

2. Its survival probability $p_i$ is calculated. A random number $r$
is generated and if $p_i<r$ the individual dies and the
process returns to step 1. Otherwise:

3. The individual moves to an adjacent site. If it is occupied,
the individual dies and the process returns to step 1. If it is
empty then:

4. If no mating partner can be found among the nearest neighbors
the individual dies and the process returns to step 1. Otherwise,
the couple produces $N$ offspring. Each of them receives a
phenotype which is the average of the parents' phenotypes, plus
a random mutation rate which can be either positive or negative. This
factor is normally smaller than 1 and different for each offspring.
In case it happens to be greater than 1, it is set equal to 1.
Similarly, if it is less than 0, it is set equal to 0.
They are placed at a distance from the parents that must
be within a radius given by the mobility parameter. If the site
is occupied, the offspring dies.

5. The parents die.

We consider an initial population of $P$
individuals with random phenotypes. They are
randomly placed in a $20\times 20$ square in the center of
the lattice, which defines an initial population
density. Then we let the population evolve
according to the above rules.

\section{Results}

In Fig. 1 we show the population density at the stationary
state as function of the mutation rate for a certain
set of parameters. These values have been calculated
by letting the initial population of $P=250$
individuals evolve and colonize
the habitat. After the population stabilizes and
stops growing, we average the density for a sufficiently
long time to obtain the maximum population density. This
maximum does not depend on the value of $P$. Instead, 
the initial population influences the survival 
probability, as will be seen later.
As is well known, the final density
depends on the value of the mutation rate and this has
an optimal value for which the density is maximum.
Given the simplicity of the model, we make no attempt
here to compare our results with real values for the
mutation rate, but rather we want to emphasize the
existence of the effect in this model and its 
dependence on other
parameters from a qualitative point of view.

Our simulations have been done by varying the
selection pressure from 1 to 3, mobility from 2 to 5
and fecundity from 6 to 8. For each combination of
these parameters, a search is made for the value
of the mutation rate which maximizes the final density.

Figure 2a shows the optimal mutation rates as function
of selection pressure for different fecundities and
a mobility value of 3. It can be seen that for
higher selection a lower mutation rate is needed
in order for the population to attain its maximum
density. This is so because at a high selection the
individuals can easily die if their fitness is not
good enough, so their phenotype must closely match that
of the landscape. Once the population has colonized the
habitat and the individuals have adjusted their phenotype
to that of the local landscape, a large mutation will
bring the individual to a point where his fitness is not
good enough to survive, so the mutation must be low. This
characteristic is independent of the number of offspring
and this can be seen in the figure, the three curves
for different fecundities almost match.

In figure 2b, the optimal mutation rate as function of
selection pressure is shown. Each curve corresponds to
a different mobility, and we have averaged the
values for different fecundities since as we mentioned above,
the dependence on fecundity is negligible. Here we note that
for higher mobility, the optimal mutation is larger.
This behavior can be explained by noting that a large
mobility means that the offspring can be put at a
position which is far from the parents from whom
they inherited their phenotype, which in the stationary
state is close to the optimal one. Since they are
now in a region where the climate is different, their
fitness is not good unless a mutation changes the
phenotype to a value that matches that of the habitat.
The larger the distance between the offspring and the
parents, the mutation that is needed will be larger.
This can be estimated noting that the change in the
optimal phenotype from site to site in the lattice of
size 200$\times$100 is 0.005. Therefore if we increase the
mobility by one, the mutation rate must increase by
that factor, and indeed the difference between the
curves in figure 2b is on the average 0.005.

We now turn attention to the behavior of the maximum
density. Figure 3 shows the maximum density
obtained at the optimal mutation rate for a mobility of 3 and
different fecundities, as function of the selection
pressure. As expected, the maximum density
increases with fecundity. Selection pressure only
plays a minor effect in lowering the maximum density
for higher selection. Since this dependence is linear,
we take an average over the different values of
selection to reduce the number of parameters. In figure
4 we can see the values of maximum density, averaged
over the selection, as function of the mobility, for
the three different values of fecundity. From these
curves we infer that the principal factor affecting
the final density is the fecundity.

Another feature that we examine is the initial condition
for the system, that is, the survival probability of the initial
population and its relation to the model parameters, in particular
the influence of the mutation rate.

We start the simulations as before, but now we let the
initial population $P$ vary from 4 to 150 individuals.
The simulation stops when all
the individuals die or when they have occupied 30\% of the
available space, when it is considered that the 
initial population survived at the specified initial density. 
For each set of the parameters,
selection pressure, fecundity and mobility
we perform $1.5\times 10^3$ independent runs in order to obtain a
survival probability for the specified set of parameters.

Figure 5 shows the results obtained from the model for
a fecundity of 6 and mobility 3. As
expected, high selection pressure lowers the survival probability
and viceversa. Similarly, the survival probability is higher
when the fecundity increases. A similar plot is obtained
in this case. In general, small populations have less chances
of survival independently of other parameters like selection,
as the figure shows. This vulnerability of small populations
is a behavior well known to biologists.
Actually one can define a {\em minimum viable population}, that is,
the smallest isolated population having a 99\% chance of remaining extant 
for 1000 years. This concept was originally introduced by Shaffer
\cite{shaff1,shaff2,rev}.

In order to better compare different sets of parameters
we reduce each of the curves shown to a single point by
averaging the survival probability over all initial
populations. Figure 6a compares the results for 
selection pressures from 1 to 3 as function of the mobility and
fecundity fixed at 7, and
Fig. 6b is analogous but the different curves correspond to
different values of fecundity at a fixed selection pressure of 1.5.  
It is found that an optimal mobility of 3 exists, which is
independent of the selection pressure and fecundity.
The existence of a maximum can be expected since a low mobility
produces overcrowding near the parents and the offspring die
because of lack of space in general. On the other hand, if the
mobility is too large, the offspring have more difficulty finding
a mating partner and also die, therefore, there should be an optimal
value in between.

Now we shall discuss the role played by the mutation
rate on these probabilities.
All the simulations above were done
at the optimal mutation rates for each set of parameters.

Depending on several factors, a population that starts to evolve
can be on benign or adverse conditions. 
By benign conditions we mean large fecundity, small selection
pressure and a small difference between the individual's
phenotype $z_i$ and the climate $F(x_i)$.

Our simulations show that if conditions are benign,
the mutation rate has a negligible effect on the
survival probability. On the contrary, if conditions
are adverse, the mutation rate does affect the fate
of the initial population.

In figure 7 we show the survival probability as a function
of the mutation rate for a population initially at the left, and
with a low fertility of 6, thus
the individuals must survive in a adverse environment. As can
be seen, the survival probability in this case is greater
for larger mutation rate, up to a maximum value and then
decreases. This can be explained in the following way:
if the conditions are adverse, the individuals must adapt (that is,
change their phenotype to match the ideal one) as soon as possible before
the whole population dies. The only way to do this is by having
a large mutation rate. Of course if the mutation is too high the
contrary effect is obtained, since harmful mutations become
more probable. In contrast, a population that develops under
conditions of high fertility and a benign habitat shows a very
weak dependence on the mutation rate.

Finally, a phase transition exists with respect to the fecundity.
In order to show that, we modified the fecundity in order to
become a continuous variable using a Poisson distribution.
In figure 8, the curves for the survival probability as function
of time for different fecundities are shown. The central curve
which is a straight line represents the critical point of the 
system, above that value of fecundity all populations survive,
and below, all populations disappear. The critical fecundity of course
depends on other parameters of the model, and therefore a
survival-extintion phase diagram can be constructed as in
\cite{pek}.

\section{Conclusions}

We have used the model of P\c ekalski \cite{pek2} to study the
influence of the mutation rate in the colonization process.
We found that, even in this simple model,
there is an optimal mutation rate for which
the final population achieves its maximum density. We also
give the dependence of this maximum with the model's parameters
such as mobility, fecundity and selection pressure. We found that
the optimal mutation rate is practically independent on the fecundity,
decreases when selection pressure increases, and increases
approximately linearly with the mobility.
Finally, we show that the initial development of the population
also depends on the mutation rate. The more adverse the living conditions
are, the higher the mutation rate must be in order for the
population to adapt quickly before it dies. We want to point out
that in this case, if the mutation rate is high and the population
survives and colonizes the habitat, it will do it with a non optimal
mutation rate, therefore the final density will be lower than the
maximum possible one. It would be interesting to study the colonization
process under the assumption of a mutation rate that is itself subject
to mutation. We are currently working on these issues.

{\em Acknowledgments}. We want to thank A.\ Bouzas and C.\ Moukarzel
for critical reading of the manuscript.

FIGURE CAPTIONS

Figure 1. Final density as function of the mutation rate using a selection
pressure of 3, mobility 2, and fecundity 6. 

Figure 2. a) Optimal mutation rate as function of selection pressure. 
The different curves correspond to fecundities of 6 (solid circles), 
7 (open circle) and 8 (asterisks). The mobility has a value of 3.
b) Optimal mutation rate as function of selection pressure. The 
lower curve corresponds to a mobility value of 2, the next one to a 
mobility of 3, and so on. Here each curve is an average of the 
curves for different fecundities.

Figure 3. Maximum population density when the whole habitat has been
colonized. The curves are for fecundity 6 (solid circles), 7 (open
circles), and 8 (open squares). The mobility is set to 3.

Figure 4. Same as figure 3, but averaged over the selection pressure
as function of the mobility.

Figure 5. Survival probability plotted as function of the initial population.
The leftmost curve is for a selection pressure of 1 and the next ones
for values of 1.5, 2, 2.5 and 3 in that order. The mobility is 3 and the
fecundity is 6.

Figure 6. a) Averaged survival probability as function of mobility for a
fecundity of 7. The uppermost curve corresponds to a selection
pressure of 1, and the next ones in descending order are for values
of 1.5, 2, 2.5 and 3.
b) Same as a) but the different curves are for
fecundities of 6 (solid circles), 7 (open circles) and
8 (asterisks). The selection pressure is fixed at 1.5.

Figure 7. Survival probability as function of the mutation rate. The 
upper curve correspponds to a population that evolves at the center of
the habitat, with a fertitlity of 8 and selection pressure 1
(benign conditions). The lower
curve has fertility 6, selection pressure 1.5 and is initially
placed at the left of the habitat (adverse conditions).

Figure 8. Survival probability of populations as function of time.
The different curves represent fecundities of 5.7, 5.705, 5.71,
5.715 and 5.72, from bottom to top. The central curve (5.71)
gives approximately the critical point of the system. A lattice
of size 800$\times$400 has been used and results are averaged over
$10^6$ realizations.

\vfill

\newpage
\begin{figure}[!t]
\begin{center}
\includegraphics[width=15cm]{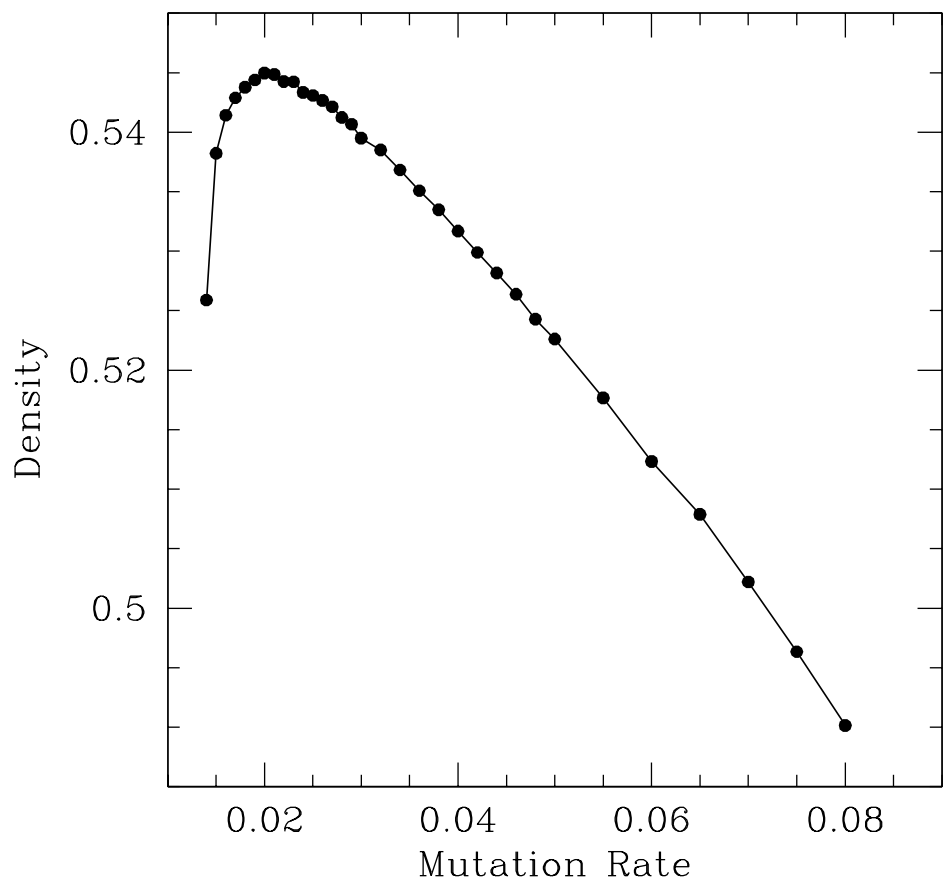}\\
\parbox{8.5cm}{\caption{
\footnotesize {R. Huerta-Quintanilla and M. Rodr\'{\i}guez-Achach}
}}
\end{center}
\end{figure}

\newpage
\begin{figure}[!t]
\begin{center}
\includegraphics[width=15cm]{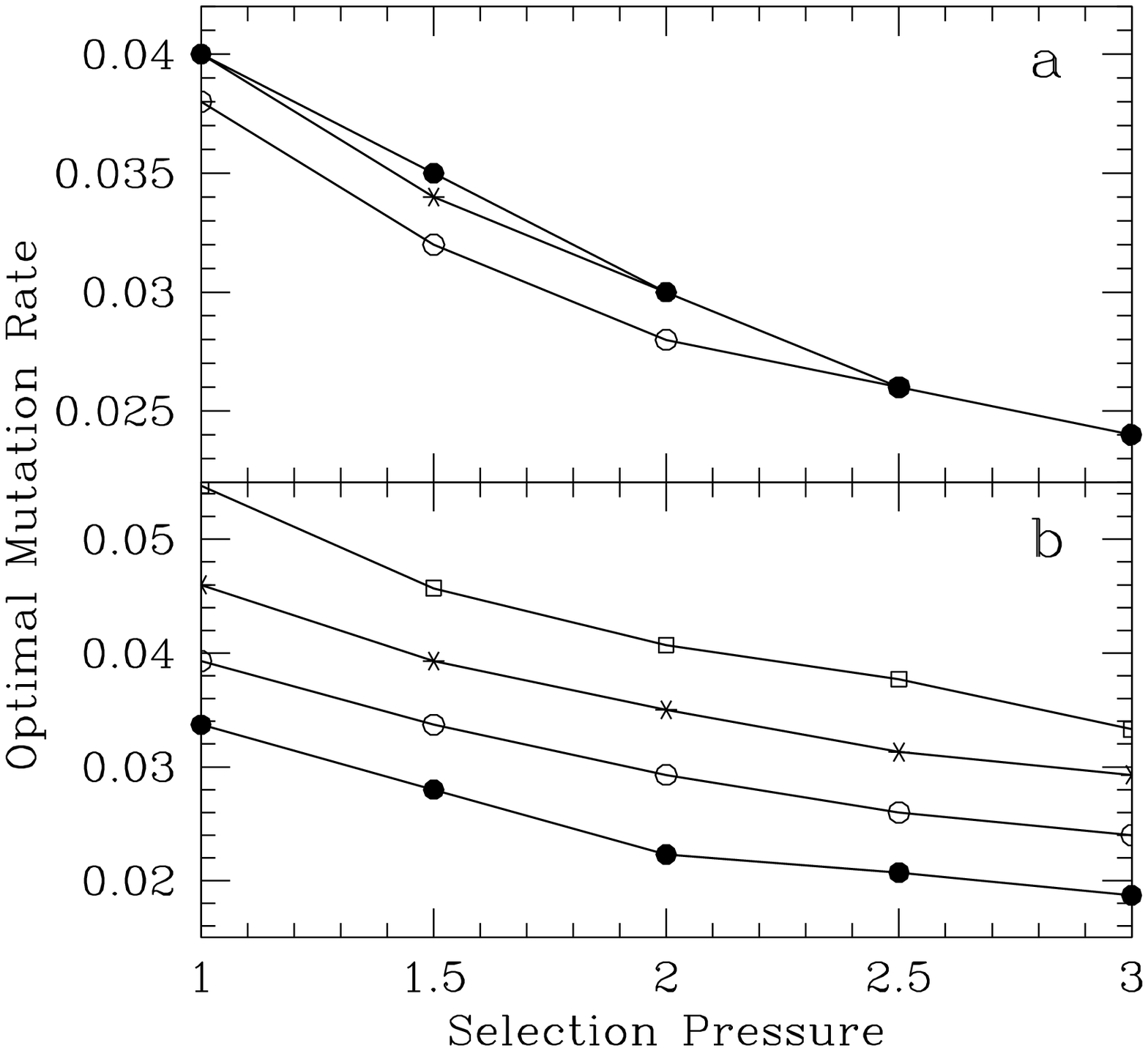}\\
\parbox{8.5cm}{\caption{
\footnotesize {R. Huerta-Quintanilla and M. Rodr\'{\i}guez-Achach}
}}
\end{center}
\end{figure}

\newpage
\begin{figure}[!t]
\begin{center}
\includegraphics[width=15cm]{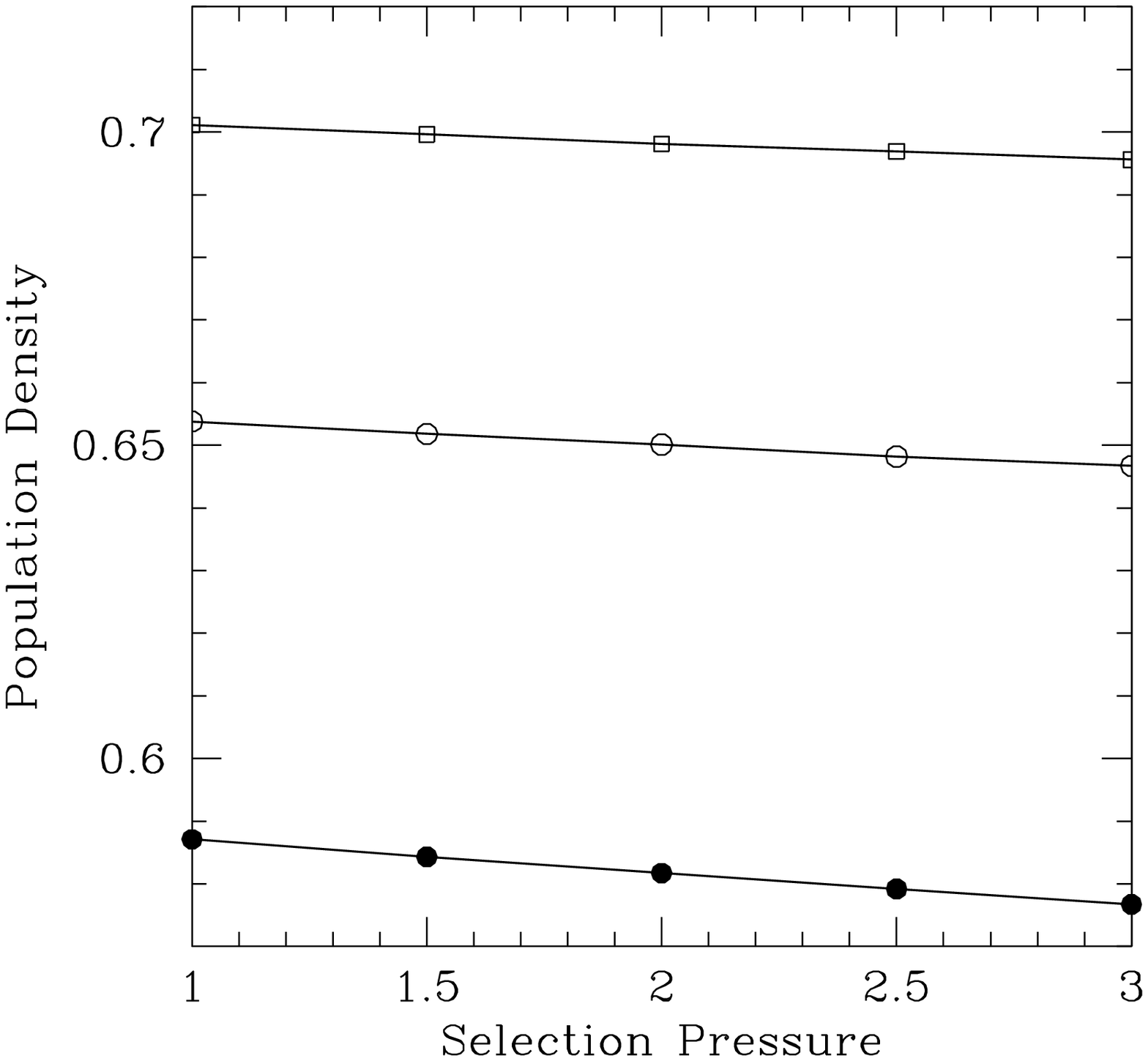}\\
\parbox{8.5cm}{\caption{
\footnotesize {R. Huerta-Quintanilla and M. Rodr\'{\i}guez-Achach}
}}
\end{center}
\end{figure}

\newpage
\begin{figure}[!t]
\begin{center}
\includegraphics[width=15cm]{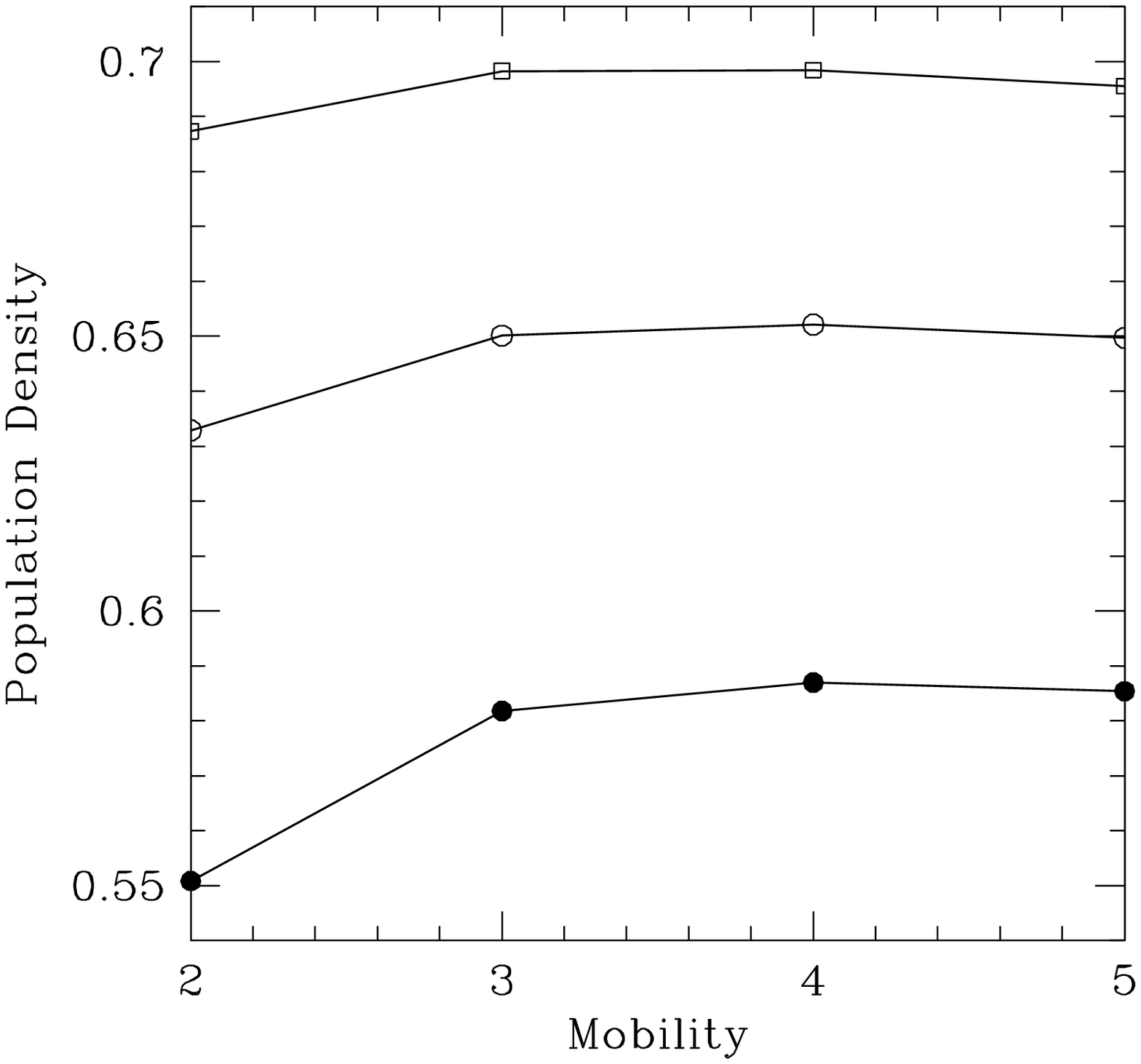}\\
\parbox{8.5cm}{\caption{
\footnotesize {R. Huerta-Quintanilla and M. Rodr\'{\i}guez-Achach}
}}
\end{center}
\end{figure}

\newpage
\begin{figure}[!t]
\begin{center}
\includegraphics[width=15cm]{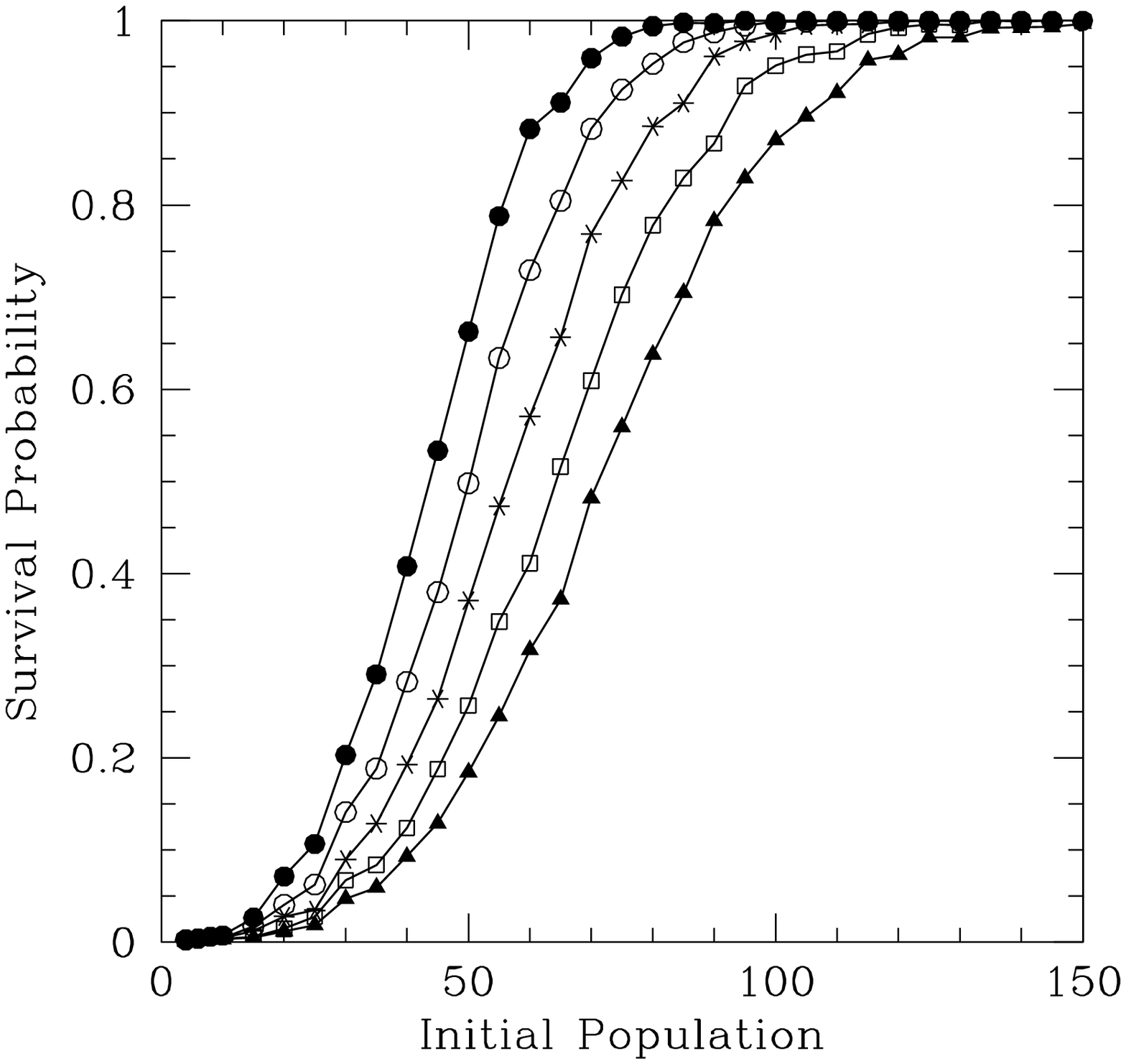}\\
\parbox{8.5cm}{\caption{
\footnotesize {R. Huerta-Quintanilla and M. Rodr\'{\i}guez-Achach}
}}
\end{center}
\end{figure}

\newpage
\begin{figure}[!t]
\begin{center}
\includegraphics[width=15cm]{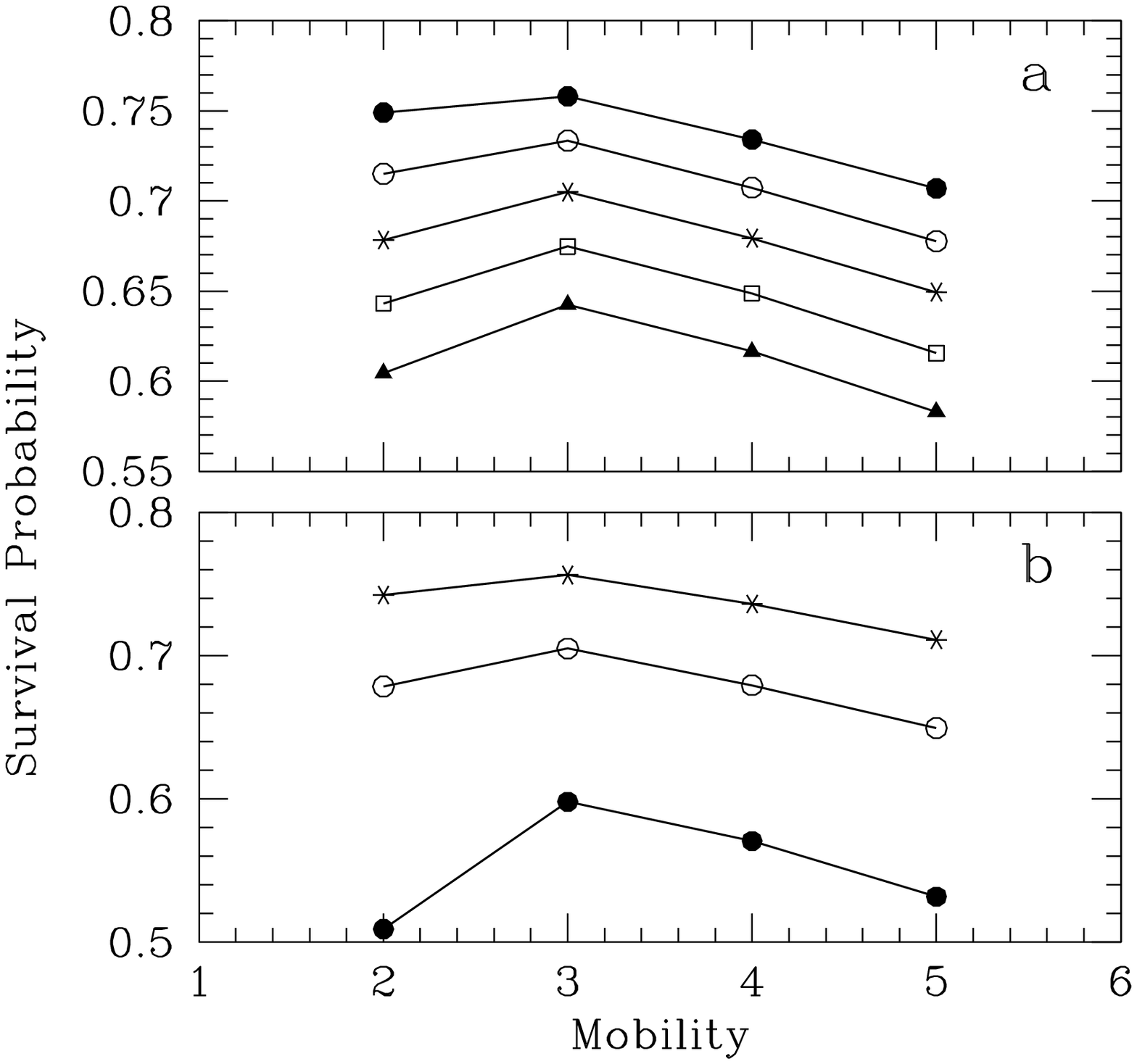}\\
\parbox{8.5cm}{\caption{
\footnotesize {R. Huerta-Quintanilla and M. Rodr\'{\i}guez-Achach}
}}
\end{center}
\end{figure}

\newpage
\begin{figure}[!t]
\begin{center}
\includegraphics[width=15cm]{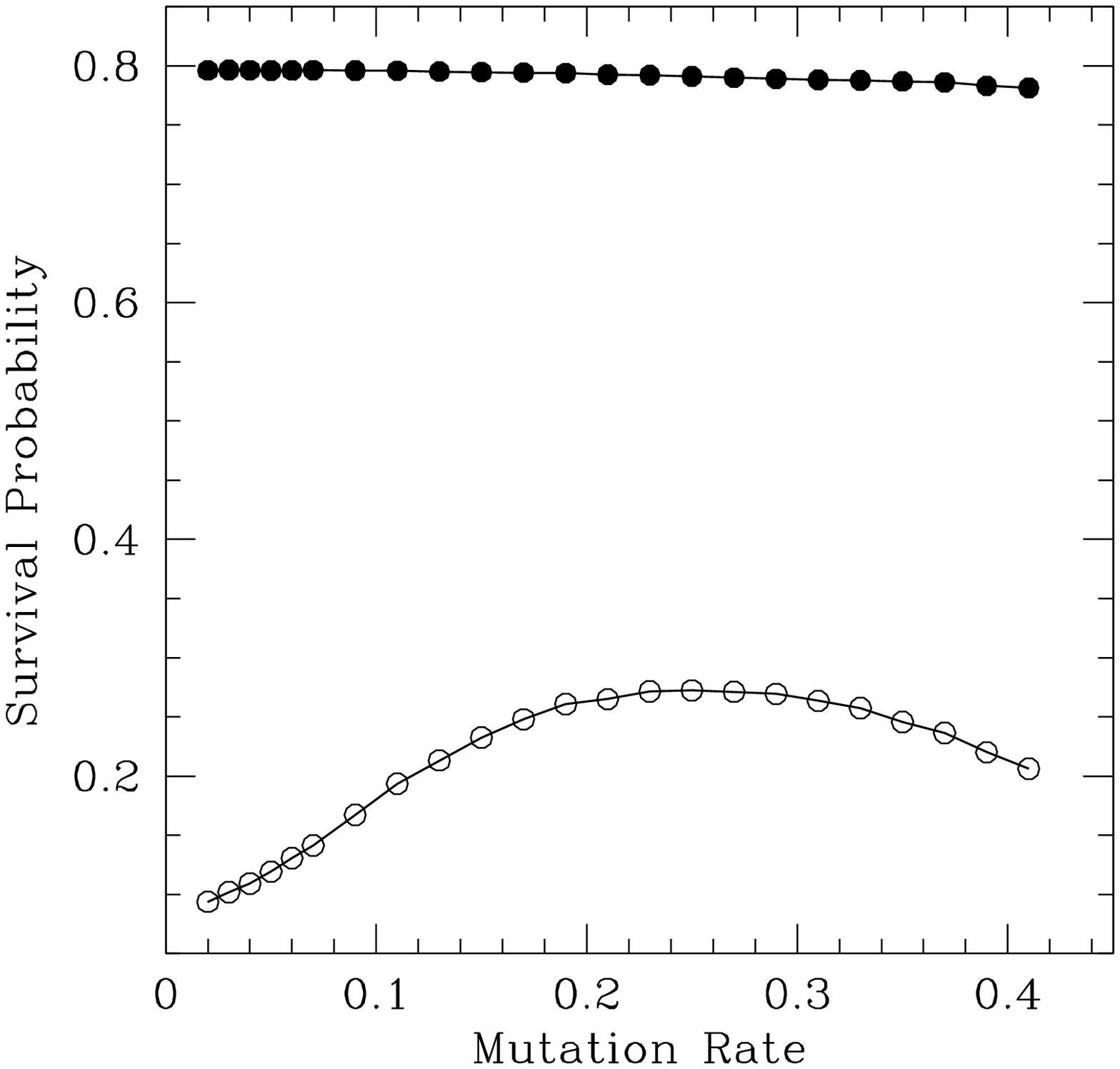}\\
\parbox{8.5cm}{\caption{
\footnotesize {R. Huerta-Quintanilla and M. Rodr\'{\i}guez-Achach}
}}
\end{center}
\end{figure}

\newpage
\begin{figure}[!t]
\begin{center}
\includegraphics[width=15cm]{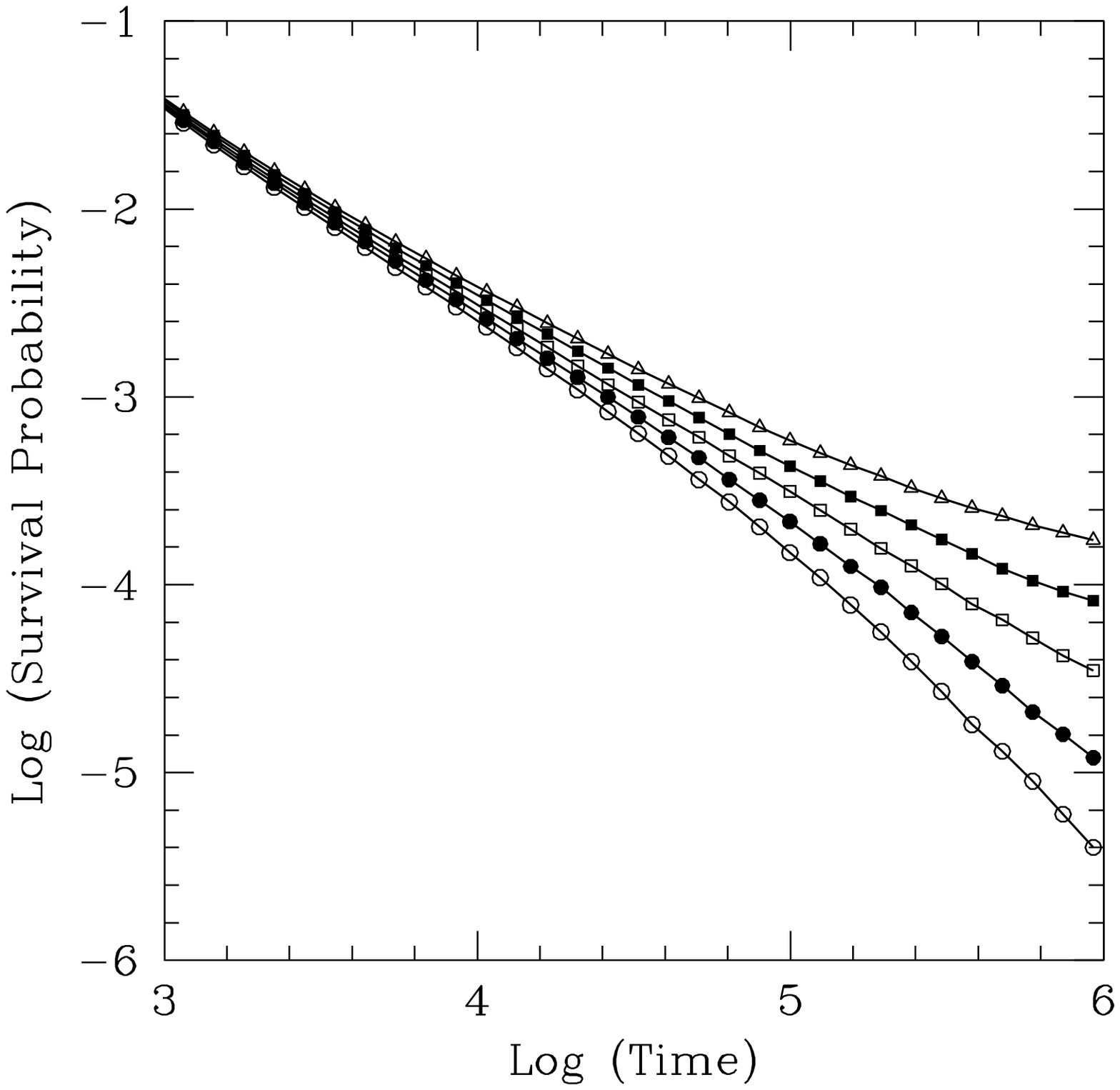}\\
\parbox{8.5cm}{\caption{
\footnotesize {R. Huerta-Quintanilla and M. Rodr\'{\i}guez-Achach}
}}
\end{center}
\end{figure}

\end{document}